\documentclass{pasj00}
\RelativeFigurePath{ps}
\draft

\begin{document}
\SetRunningHead{Author(s) in page-head}{Running Head}
\Received{0}
\Accepted{0}

\title{Supernova Remnants in the AKARI IRC Survey of the Large Magellanic Cloud}

\author{Ji Yeon \textsc{Seok} \altaffilmark{1},
Bon-Chul \textsc{Koo}\altaffilmark{1},
Takashi \textsc{Onaka}\altaffilmark{2},
Yoshifusa \textsc{Ita}\altaffilmark{3},
Ho-Gyu \textsc{Lee}\altaffilmark{1}, \\
Jae-Joon \textsc{Lee}\altaffilmark{4},
Dae-Sik \textsc{Moon}\altaffilmark{5},
Itsuki \textsc{Sakon}\altaffilmark{2},
Hidehiro \textsc{Kaneda}\altaffilmark{6},
Hyung Mok \textsc{Lee}\altaffilmark{1},
Myung Gyoon \textsc{Lee}\altaffilmark{1},
Sung Eun \textsc{Kim}\altaffilmark{7}
}
\altaffiltext{1}{Department of Physics and Astronomy, 
	Seoul National University, Seoul 151-742, KOREA}
\email{jyseok@astro.snu.ac.kr; koo@astrohi.snu.ac.kr; hglee@astro.snu.ac.kr;
	 hmlee@astro.snu.ac.kr; mglee@astro.snu.ac.kr}
\altaffiltext{2}{Department of Astronomy, Graduate School of Science,
University of Tokyo, Bunkyo-ku, Tokyo 113-0003, JAPAN}
\email{onaka@astron.s.u-tokyo.ac.jp; isakon@maitta.astron.s.u-tokyo.ac.jp}
\altaffiltext{3}{National Astronomical Observatory of Japan, 2-21-1 Osawa, Mitaka, Tokyo 181-8588, JAPAN}
\email{yoshifusa.ita@nao.ac.jp}
\altaffiltext{4}{Department of Astronomy and Astrophysics, Pennsylvania State University, 525 Davey Laboratory, University Park, PA 16802, USA}
\email{lee@astro.psu.edu}
\altaffiltext{5}{Department of Astronomy and Astrophysics, University of Toronto, Toronto, ON M5S 3H4, CANADA}
\email{moon@astro.utoronto.ca} 
\altaffiltext{6}{Institute of Space and Astronautical Science, Japan Aerospace Exploration Agency, Sagamihara, Kanagawa 229-8510, JAPAN}
\email{kaneda@ir.isas.jaxa.jp} 
\altaffiltext{7}{Department of Astronomy and Space Science, Sejong University, Seoul 143-747, KOREA}
\email{sek@sejong.ac.kr}


%


\maketitle

\begin{abstract}
We present the near- to mid-infared study of supernova remnants (SNRs) using the AKARI IRC Survey of the Large Magellanic Cloud (LMC). The LMC survey observed about a 10 square degree area of the LMC in five bands centered at 3, 7, 11, 15, and 24 \micron~using the Infrared Camera (IRC) aboard AKARI. The number of SNRs in the survey area is 21, which is about a half of the known LMC SNRs. We systematically examined the AKARI images and identified eight SNRs with distinguishable infrared emission. All of them were detected at $\gtrsim 10$ \micron~and some at 3 and 7 \micron, too. We present their AKARI images and fluxes. In the 11/15 \micron~versus 15/24 \micron~color-color diagram, the SNRs appear to be aligned along a modified blackbody curve, representing thermal emission from dust at temperatures between 90 and 190 K. There is a good correlation between the 24 \micron~and X-ray fluxes of the SNRs. It was also found that there is a good correlation between the 24 \micron~and radio fluxes even if there is no direct physical connection between them. We considered the origin of the detected mid-infrared emission in individual SNRs. We conclude that the mid-infrared emissions in five SNRs that show morphologies similar to the X-rays are dominated by thermal emission from hot dust heated by X-ray emitting plasma. Their 15/24 \micron~color temperatures are generally higher than the Spitzer 24/70 \micron~color temperatures, which suggests that a single-temperature dust model cannot describe the full spectral energy distribution (SED) of the SNRs. It also implies that our understanding of the full SED is essential for estimating the dust destruction rate of grains by SNR shocks.

\end{abstract}

\section{Introduction}

A supernova (SN) explosion is one of the most energetic events in the Universe, ejecting various elements of stellar mass with enormous energy. It plays an important role in the evolution of the interstellar medium (ISM) by generating strong shocks, which heat and accelerate the medium and destroy dust grains. A significant amount of dust grains may also be formed in the SN ejecta. These processes can be studied by observing the remnants of the explosion, supernova remnants (SNRs), at various wavelengths. In particular, infrared (IR) observation can significantly improve our understanding on physical processes associated with dust grains because they, either newly synthesized or swept-up by SN shocks, radiate essentially only in the IR. It is also useful to explore the nature of SNRs and their environments through atomic, ionic, and molecular lines in the IR, and also sometimes through synchrotron emission from relativistic electrons.

The Large Magellanic Cloud (LMC) contains more than forty radio/X-ray SNRs, and offers a unique opportunity for studying SNRs, owing to its location. At a distance of 51.4 kpc \citep{panagia99} far off from the galactic plane, we can see the detailed structure of SNRs without much confusion by foreground or background material. In spite of such advantages and the importance as our neighbor galaxy, IR studies on SNRs in the LMC have not flourished compared to radio or X-ray due to observational obstacles. The Infrared Astronomy Satellite (IRAS) opened an era of IR studies in 1980s by detecting IR emission from a number of SNRs \citep{graham87}. The instrumental ability such as spatial resolution and wavelength coverage, however, was not sufficient to reveal the nature of the IR emission from individual SNRs.

The Spitzer Space Telescope, launched in 2003, has made significant  progress in this field. Using the recent Spitzer Survey of the LMC: Surveying the Agents of a Galaxy Evolution (SAGE; \cite{meixner06}), it became possible to investigate almost whole region of the LMC (7 $\times$ 7 $\mathrm{deg}^{2}$) at $3-8~\micron$ and at $24-160~\micron$. In addition, a separate survey of 39 LMC SNRs was conducted to study the interstellar dust lifecycle in terms of ejecta formation by SNe and dust destruction by SNR blast waves (PI: K. Borkowski). Its preliminary results suggest that a substantial amount of small grains are destroyed in both Type Ia \citep{borko06} and core-collapse SNRs \citep{b_willi06}. \citet{williams06} carried out Spitzer observations of six SNRs in the LMC and found line emission as a significant contributor to the IR emission. \citet{tappe06} reported on the detection of bright mid-infrared (MIR) emission together with the polycyclic aromatic hydrocarbon (PAH) features in the oxygen-rich SNR N132D using Spitzer imaging and spectroscopy.   

Recently, the AKARI infrared space telescope, launched on 2006 February 21 (UT), has performed a large scale survey of the LMC (\cite{ita08}; see \S~\ref{sec:lslmc}). AKARI has continuous coverage of imaging from $2.5-26~\micron$, which is a powerful tool to investigate the IR trait of SNRs. In particular, the 11 and 15 $\micron$ bands are unique to AKARI, and could provide important information unobtainable with the Spitzer observations. In this paper, we report on the detection of IR emission in eight out of 21 SNRs in the AKARI IRC survey field. We present their images and show that there is a good correlation between the MIR and X-ray/radio fluxes. We discuss the origin of their MIR emission.

\begin{figure}
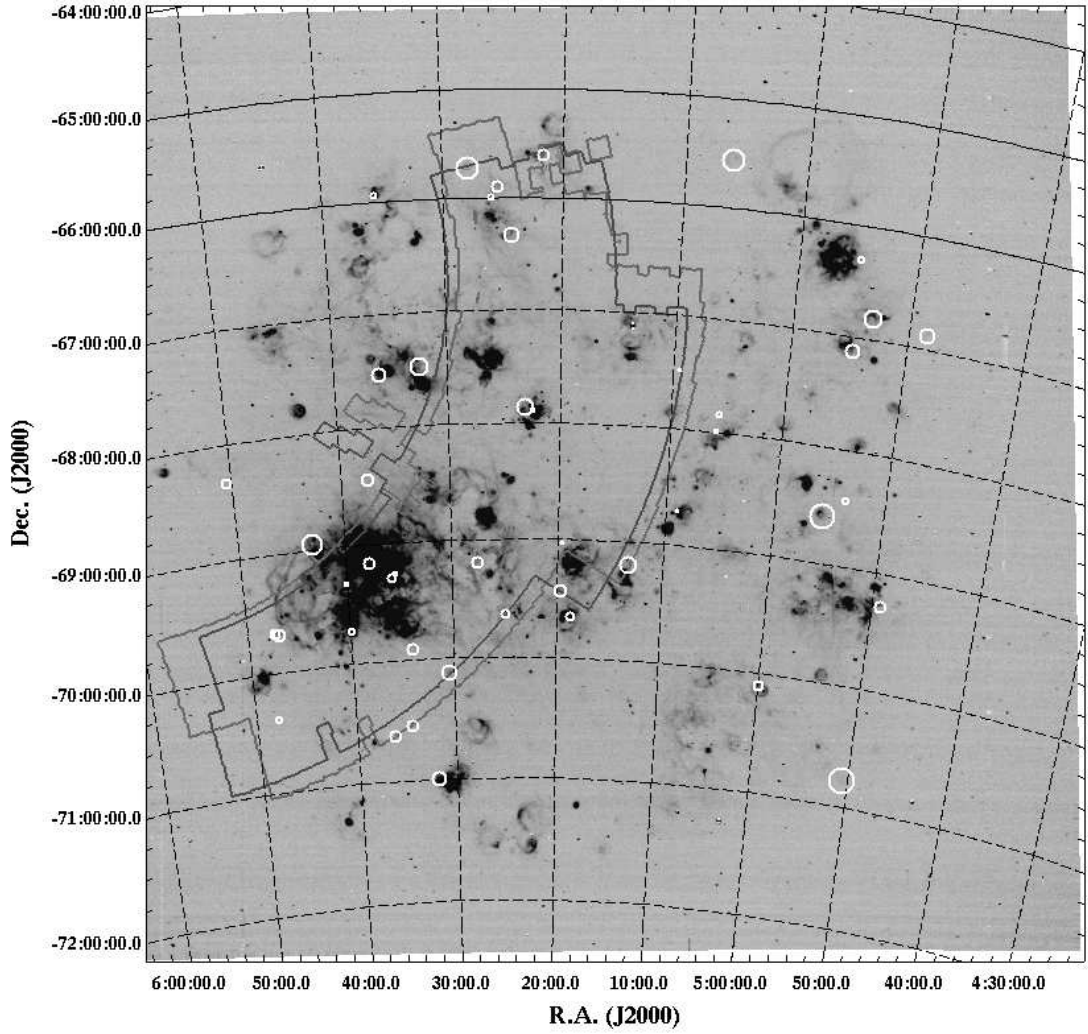

  \begin{center}
    \FigureFile(150mm,150mm){egrtrHaLMCsurveyarea.eps}
  \end{center}
  \caption{Observed area of the AKARI IRC survey, overlaid on the H$\alpha$ image of the Large Magellanic Cloud~\citep{kim98}. The dark gray and the light-gray lines indicate the survey area in the NIR/MIR-S and the MIR-L bands, respectively. Note that there is a discrepancy of the observed areas among the bands due to their separate field-of-views. The marked circles represent the location of the known SNRs and SNR candidates, and the size of each circle is proportional to that of the corresponding SNR.}
  \label{fig:lmcHa}
\end{figure}

\section{AKARI LMC survey and SNR identification} \label{sec:lslmc}

The data we used is from the AKARI large-scale survey of the Large Magellanic Cloud (PI: T. Onaka), one of the three AKARI large-scale survey programs. The survey was performed from May 2006 to July 2007 using the on-board instrument Infrared Camera (IRC; \cite{onaka07}). IRC has three channels: NIR/MIR-S channels sharing the same $10\arcmin \times 10\arcmin$ field-of-view and MIR-L channel observing the sky about $25\arcmin$ away from the NIR/MIR-S field-of-view. The survey covers about a 10 $\mathrm{deg}^{2} $ region of the LMC, which includes most of the major regions of the LMC in all three channels (figure~\ref{fig:lmcHa}). Areas near the boundary of this targeted area were covered in either NIR/MIR-S or MIR-L channel. The imaging observations were carried out in five bands: {\it N3} ($2.7-3.8 ~\micron $), {\it S7} ($5.9-8.4 ~\micron $ ), {\it S11} ($8.5-13.1 ~\micron $), {\it L15} ($12.6-19.4 ~\micron $), and {\it L24} ($20.3-26.5 ~\micron $). The total integration time was 133 s for an {\it N3} band image and 147 s for the other band images \citep{ita08}, and the $ 5 \sigma $ sensitivities per poiniting were 16, 74, 132, 279, and 584 $\mu$Jy in {\it N3}, {\it S7}, {\it S11}, {\it L15}, and {\it L24}, respectively \citep{onaka07}. We processed the images by using the standard IRC imaging data reduction pipeline \citep{irc_idum}. In addition to these imaging observations, near-infrared (NIR) low-resolution slit-less spectroscopy ($R\sim20$) was performed for the same 10 $ \mathrm{deg}^{2}$ region in $2-5 ~\micron$ with the NIR prism spectroscopic mode (NP) of IRC. We also examined the spectroscopic data of individual SNRs. However, since most of them are not visible in this spectral range, and even the SNRs showing NIR spectra have difficulties to extract useful information due to confusion from background and/or nearby sources, we do not mention the NP results in this paper.  

\begin{figure}
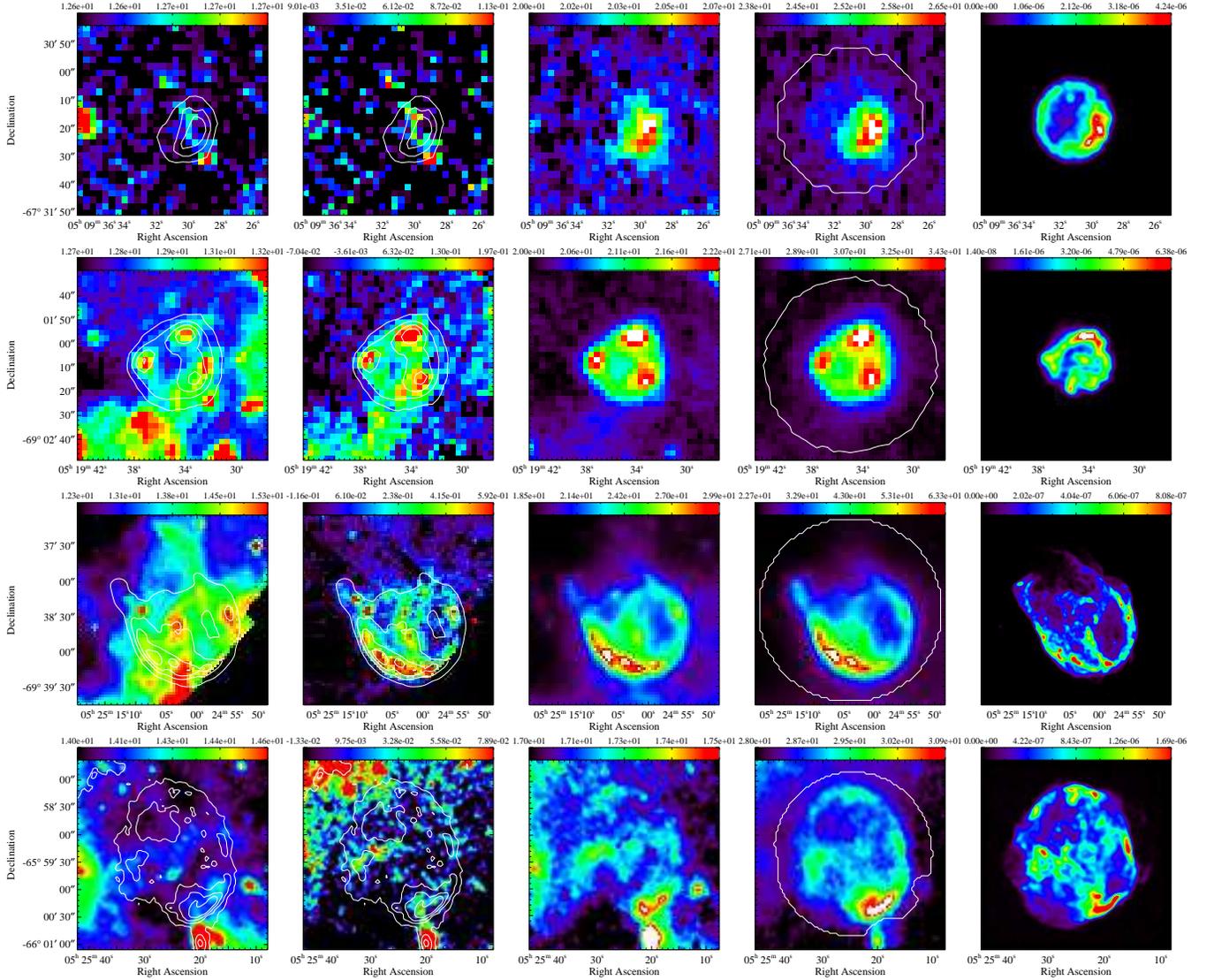

  \begin{center}
    \FigureFile(180mm,200mm){figure2.eps}
    \end{center}
  \caption{From top to bottm: AKARI images of 0509--67.5, 0519--69.0, N132D, and N49B. For each SNR, {\it S11}, ``{\it S11}-{\it S7}'', {\it L15}, and {\it L24} images are shown together with their Chandra X-ray (0.3--2.1 keV) images for comparison. {\it S11}-{\it S7} images are made by subtracting the scaled {\it S7} images from the {\it S11} images (see text for details). The Chandra images are from the CSRC. The contours in the {\it S11} and {\it S11}-{\it S7} images show the brightness distribution of SNRs in the {\it L24} band, and the contours in the {\it L24} images represent the area used for flux derivations. For N49B, the images are smoothed with a two-pixel Gaussian. The units on the colorbar of the AKARI images are MJy/sr, and the Chandra are $\mathrm{counts/cm^{2}/s}$.}
  \label{fig:mirsnr1}
\end{figure}

\begin{figure}
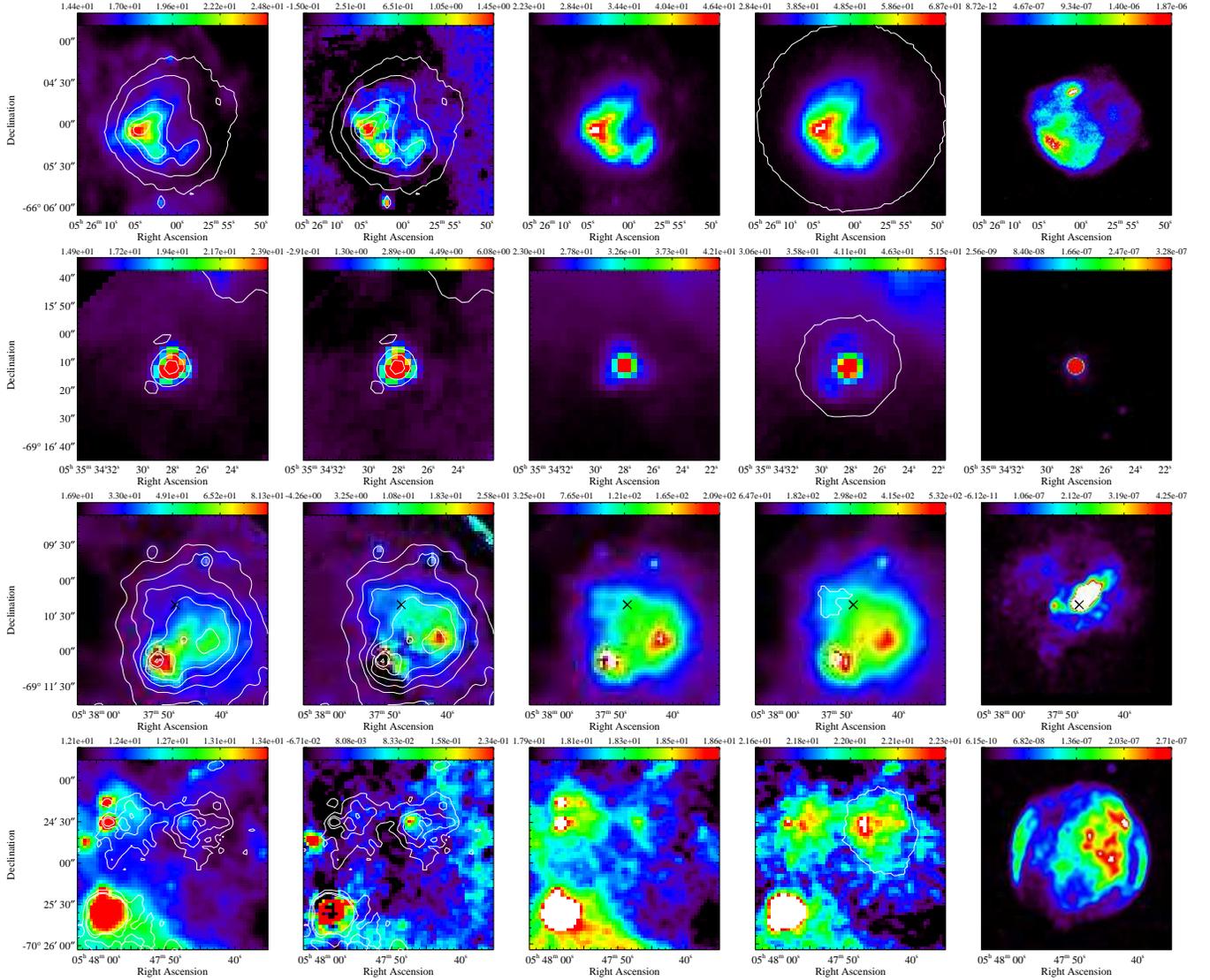

  \begin{center}
    \FigureFile(180mm,200mm){figure3.eps}
    \end{center}
  \caption{From top to bottm: Same as figure~\ref{fig:mirsnr1} but for N49, SN 1987A, N157B, and 0548--70.4.
``$\times$'' marks on the third row represent the position of pulsar in N157B. The images of 0548--70.4 are smoothed with a two-pixel Gaussian. Note that for N157B and 0548--70.4, the flux has been extracted from the limited region due to the confusion (see \S~\ref{sec:mfc}).}
  \label{fig:mirsnr2}
\end{figure}

Twenty one, previously-known LMC SNRs are included in the AKARI LMC survey area. Figure~\ref{fig:lmcHa} shows their positions on the H$\alpha$ image obtained by \citet{kim98} together with the boundary of the area covered in the AKARI LMC survey. We examined whether there is IR emission associated with the SNRs. Some SNRs are embedded within IR-bright HII regions or HII complexes such as 30 Doradus, so that a careful inspection was required to discriminate the SNR emission from the surrounding medium. The reference data we used for confirming our SNR identification were the X-ray images from the Chandra Supernova Remnant Catalog (CSRC) page\footnote{http://hea-www.harvard.edu/ChandraSNR/}, the radio images from 4.8 GHz survey using the Australia Telescope Compact Array (ATCA; \cite{dickel05}), and the optical images from the Magellanic Cloud Emission Line Survey (MCELS; \cite{smith00}). Table~\ref{tab:lmcsnr} lists 21 SNRs in the survey area, and summarizes the results of the search. We have identified eight SNRs in total, which have associated IR emission in the NIR and/or MIR bands including three Type Ia SNRs and five core-collapse SNRs (Type II SNRs for simplification, hereafter). Most of the identified SNRs have distinct shell-like features in the MIR-L bands as well as some related emission in the {\it S11} band. Only a few SNRs show any emission possibly associated with the SNR in shorter wave bands. Figures~\ref{fig:mirsnr1} and \ref{fig:mirsnr2} show the AKARI MIR-band images of the identified eight SNRs, and figure \ref{fig:nirsnr}~shows the AKARI {\it N3} and {\it S7} band images of the two SNRs (excluding SN 1987A) with the related NIR features. 
   
\section{Infrared Properties of SNRs} \label{sec:irprop_snr}
\subsection{Morphologies and Fluxes} \label{sec:mfc}

 Most of the identified SNRs are visible in the {\it S11}, {\it L15}, and {\it L24} bands. In the {\it S11} band, however, the SNR features are confused by the foreground and background emission which might be dominated by strong PAH 11.3 $\micron$ band emission from dust. We subtracted this background emission using the {\it S7} band image which is also dominated by the PAH emission from the background. We estimated the mean ratio of {\it S11} to {\it S7} surface brightness of the background emission in each SNR field from a pixel-to-pixel plot of {\it S11} versus {\it S7} brightness and subtract the scaled {\it S7} image from the {\it S11} image, i.e., {\it S11}$-(a\times${\it S7} $+b$), where the scaling factors a and b range $0.4 -1.4$~and $9-14$~MJy/sr, respectively. The resulting ``{\it S11}-{\it S7}'' image shows the SNR features much more clearly than the original {\it S11} image. In figures \ref{fig:mirsnr1} and \ref{fig:mirsnr2}, we show the {\it S11}, {\it S11}-{\it S7}, {\it L15}, and {\it L24} images of eight identified SNRs together with their Chandra X-ray images.

\begin{figure}
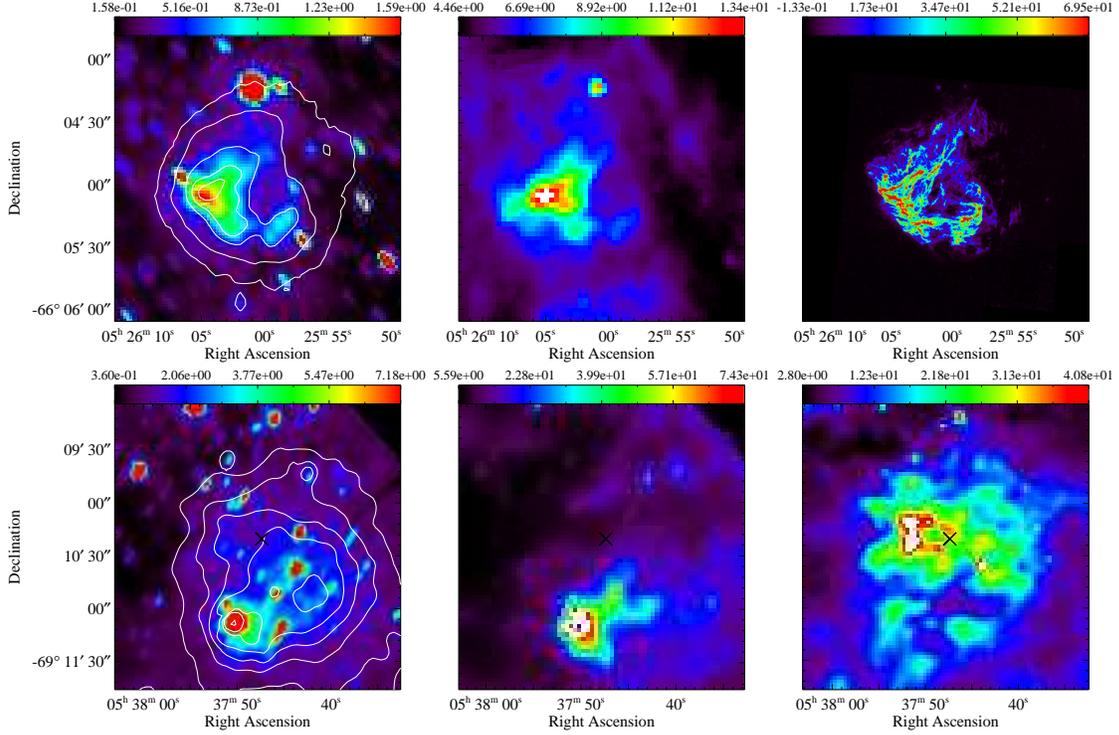

  \begin{center}
    \FigureFile(150mm,150mm){figure4.eps}
    \end{center}
  \caption{{\it N3}, {\it S7} and H$\alpha$ images of N49 (top) and N157B (bottom). These two (and SN 1987A) are SNRs identified in these bands. The contours in the {\it N3} images show the brightness distribution of SNRs in the {\it L24} band. Optical images of N49 and N157B are from the Hubble Space Telescope (HST) WFPC2 \citep{bilikova07} and the CTIO Curtis Schmidt Telescope\footnotemark, respectively.}  
   \label{fig:nirsnr}
\end{figure}
\footnotetext{Preliminary MCELS data are available from the homepage of MCELS, http://www.ctio.noao.edu/$\sim$mcels/.} 

In all of the sources, we can see distinct MIR emission features at the position of SNRs in the {\it S11}, {\it L15}, and {\it L24} bands. The first five sources (0509--67.5, 0519--69.0, N132D, N49B, and N49) show shell-like structures that are similar to the X-ray SNRs, so that the association of the IR emission with the SNRs is conclusive. The MIR brightness distribution appears to be correlated with the X-ray brightness distribution in general, but in 0519--69.0 and N49, the morphologies are somewhat different, e.g., their peak positions do not coincide. SN 1987A is not resolved in our observation, but the positional coincidence verifies the association. N157B is a Crab-like SNR and there is {\it no} detectable MIR emission associated with the X-ray/radio pulsar wind nebula (PWN)\footnotemark.\footnotetext{If we extrapolate the radio spectrum of the PWN \citep{lazendic00}, the expected surface brightness at the radio peak, ($\alpha$, $\delta$) = ($05^{h}37^{m}45^{s}, -69^{\circ}10'11''$), is estimated to be 2.7 and 3.0 MJy/sr at 7 and 11 \micron, which is greater than the 2 $\sigma$~detection limit, 0.7 and 1.9 MJy/sr, in the {\it S7} and {\it S11} bands, respectively. This suggests that the PWN has a spectral break at a wavelength longer than 11 $\micron$.} (The very bright emission to the south of the remnant is associated with star-forming regions, not with the SNR.) But this remnant has bright, extended H$\alpha$ emission with a strong peak to the east of the PWN \citep{chu92}, and a faint `horseshoe-shaped' structure near the center of the field corresponds to the bright optical filaments (see next); 0548--70.4 is located in a rather complicated field with several MIR sources around, but still the MIR emission corresponding to the X-ray bright interior can be clearly seen. The brightness distribution of the MIR emission, however, appears to be different from that of the X-rays.The limbs are barely visible only in {\it L24} band. For all SNRs in figures \ref{fig:mirsnr1} and \ref{fig:mirsnr2}, Spitzer obtained 24 $\micron$ images \citep{borko06, b_willi06, williams06}, and our {\it L24} images are consistent with them. The AKARI {\it S11} and {\it L15} images are new, and the fact that each SNR shows compatible morphologies in three bands suggests that the MIR emissions in these three bands are of the same origin.

In shorter wavebands, three SNRs are clearly visible: N49, SN 1987A, and N157B. N132D, the MIR brightest SNR does not show any distinct NIR emission corresponding to the features seen in the MIR bands, which is partly due to its location in the complex area (cf. \cite{tappe06}). The {\it N3} and {\it S7} images of N49 and N157B are shown in figure \ref{fig:nirsnr} along with their H$\alpha$ images. N49 has a bright wedge-shaped feature in the eastern part of the remnant, which matches well with the bright optical filaments. N157B is contaminated by emission from the bright source below the remnant, but still we can identify the features corresponding to the optical filamentary structures, including the bright horseshoe-shaped one. The optical counterparts can be distinguished in all IRC bands except for {\it S7}. These {\it N3} and {\it S7} images are consistent with the Spitzer images, which have a bit higher resolution \citep{williams06}. We briefly describe the characteristics of each source in the Appendix.  
       
For the identified SNRs, we derived their fluxes in each band (table~\ref{tab:ircflux}). The {\it S11} fluxes were derived from the {\it S11}-{\it S7} images, except N49 and SN 1987A, which show appreciable emission in the {\it S7} band. The areas used for flux estimation are marked in the {\it L24} images of figure \ref{fig:mirsnr1} and \ref{fig:mirsnr2}. The fluxes are extracted from the entire SNR area, except for three SNRs: N132D, N157B, and 0548--70.4. N132D is incompletely covered in the NIR and MIR-S bands, so that we could obtain the total fluxes only in the MIR-L. Because the fraction of the uncovered area, however, is small, we estimated the total {\it S11} flux by first deriving the ratio of the {\it S11} to {\it L15} flux from the shared area and then by multiplying it to the total {\it L15} flux. 
N157B is confused by the southern star-forming regions, so we extract only the fluxes from the horseshoe-shape region in order to avoid any contamination from other sources. 0548--70.4 is located in a complicated area, and the background stars are located near the eastern rim. 

Our measured IR fluxes are uncertain by  $10-30~\%$~($1~\sigma$) considering the uncertainties in measurements, background estimation, and the absolute calibration. For most sources, the error in the absolute calibration dominates, which is about 10\% (see \cite{koo07}). The IRC fluxes are those at reference wavelengths, assuming a flux distribution of $f_{\lambda} \propto \lambda^{-1}$ \citep{irc_idum}. We did not apply any color-correction since the origin of the IR emission can differ from source to source. All flux values that we employed for further analysis were not the color-corrected ones, except those described in \S~\ref{sec:irorigin}, where we consider the dust properties. The derived IRC fluxes are mostly consistent with the published Spitzer fluxes (e.g., \cite{borko06}). For SN 1987A, \citet{dwek08} showed that its flux at 24 \micron~was $\sim30$ mJy on February 4, 2004 (day 6190) and increased by a factor of 2 after 947 days (day 7137). The AKARI spectrum was obtained from 2006 October 31 to November 4 (day $7190-7194$), and the derived IRC fluxes are similar to those of the latest spectra with the Spitzer Infrared Spectrograph (IRS). 

\begin{figure}
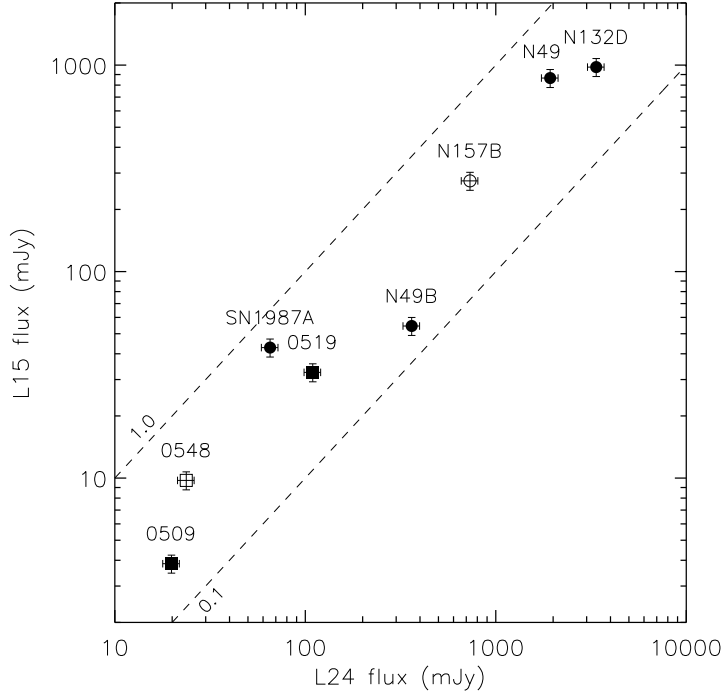

  \begin{center}
    \FigureFile(100mm,100mm){L15vsL24.eps}
    \end{center}
  \caption{{\it L15} flux versus {\it L24} flux for eight identified SNRs. Square symbols are for Type Ia SNRs and circles for Type II SNRs. Open symbols are for the objects of which fluxes were extracted from limited areas. We use the same symbols hereafter for all plots. The dashed lines show {\it L15}/{\it L24} = 0.1 and 1.}
  \label{fig:c15oc24}
\end{figure}

We compare the {\it L15} and {\it L24} fluxes of the identified SNRs in figure~\ref{fig:c15oc24}. There is a tight correlation between the two fluxes, as expected, but the {\it L15}/{\it L24} flux ratio ranges from 0.15 up to 0.7. The brightest SNRs are N132D and N49. These two SNRs are interacting with their ambient molecular clouds \citep{banas97}, and the bright IR emission may be related to the interaction. Type Ia SNRs are the faintest among the identified SNRs with ({\it L24}, {\it L15}) = (20, 3.9) to (110, 33) (mJy), while Type II SNRs are located in the upper right part, where ({\it L24}, {\it L15}) = (65, 43) to (3370, 980) (mJy). 

\begin{figure}
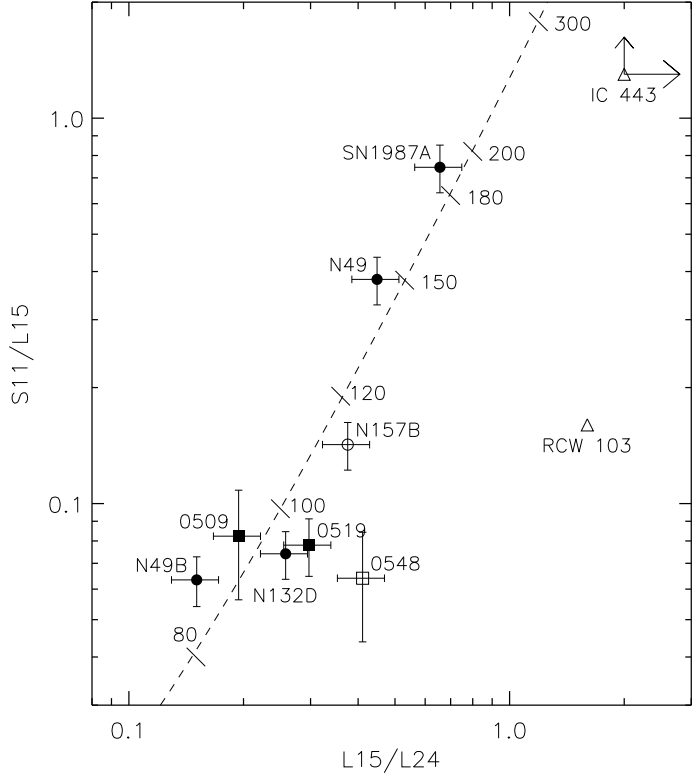

  \begin{center}
    \FigureFile(100mm,100mm){lmcsnrCCD.eps}
    \end{center}
  \caption{{\it S11}/{\it L15} flux ratio versus {\it L15}/{\it L24} flux ratio. The dashed line represents the expected ratios from modified blackbody curve of thermal dust emission. Dust temperatures are marked along the line. We also show the synthetic colors of two Galactic SNRs, RCW 103 and IC 443 with triangles (see text for an explanation).}
  \label{fig:snrccd}
\end{figure}

\subsection{IRC Colors} \label{sec:irccolor}

Figure \ref{fig:snrccd} is a ``color-color'' diagram comparing the {\it S11}/{\it L15} to {\it L15}/{\it L24} flux ratios. In the figure, the dashed line represents the relation between the flux ratios for thermal dust emission. The dust emission model that we adopt is a single-temperature modified blackbody, where the flux density, $F_{\nu}$, is given by 
\begin{equation}
   F_{\nu} = \frac{\kappa_{\nu} B_{\nu}(T_{d})}{ d^{2}} M_{d}    
   \label{eq:modbb}
\end{equation}
where $T_{d}$ is the dust temperature, $M_{d}$ is the dust mass, $\kappa_{\nu}$ is the dust mass absorption coefficient, $B_{\nu}(T_{d})$ the Planck function, and $d$ the distance to the LMC, taken to be $d=51.4$ kpc \citep{panagia99}. For the absorption coefficient, we adopt the ``average'' LMC model of \citet{wdraine01} which consists of mixture of carbonaceous grains and amorphous silicate grains with the maximum carbon abundance being in very small grain population\footnote{Absorption coefficient data files are from the homepage of Bruce T. Draine, available at http://www.astro.princeton.edu/$\sim$draine.}. We also show the synthetic colors of two well-known Galactic SNRs, RCW 103 and IC 443, for a comparison. RCW 103 is a young ($\sim 10^3$~yr) SNR with fast ($\gtrsim 300$~km/s) shocks interacting with dense CSM/ISM, and its MIR spectrum is dominated by forbidden lines from Ar, Ne, O, and Fe ions \citep{oli99}. On the other hand, IC 443 is a prototype of old SNRs interacting with molecular clouds, and its MIR spectrum is dominated by pure rotational H$_2$ lines \citep{neu07}. We synthesized their IRC colors from published spectra to obtain ({\it L15}/{\it L24}, {\it S11}/{\it L15}) = (1.6, 0.16) and (90, 4.3) for RCW 103 and IC 443, respectively. 

Figure \ref{fig:snrccd} shows that the MIR flux ratios of the AKARI LMC 
SNRs are quite different from the two line-dominated Galactic SNRs, and 
well aligned along the dust-emission line. Note that Type Ia SNRs are located in the lower left part and SN 1987A has the highest ratios. This good alignment seems to suggest that the MIR emission from these SNRs is, or at least dominated by, thermal dust emission. Indeed, for N132D and SN 1987A, the Spitzer spectroscopic observations showed that the contribution from ionic/molecular lines or PAH emission is small in these SNRs \citep{tappe06, bouchet06}. Meanwhile, the MIR emission of N49, which is also lying close to the dust-emission line, has been found to be dominated by ionic lines from shocked gas \citep{williams06}. However, it is not likely that this mature ($\sim 6600$~yr) remnant would have a sufficiently hugh dust temperature ($\sim 150$~K) to lie along its position on the dust-emission line. Therefore, although the fact that an SNR is located close to the dust-emission line in figure \ref{fig:snrccd} alone does not assure that its MIR emission is thermal dust continuum emission, the diagram is still helpful to distinguish between line-dominated and dust-dominated SNRs, taking account of their physical contexts. We discuss the origin of the observed MIR emissions in \S~5.

\section{Comparison with Other Wavebands}\label{sec:irxrra}
\subsection{IR vs. X-ray}\label{sec:irxr}

\begin{figure}
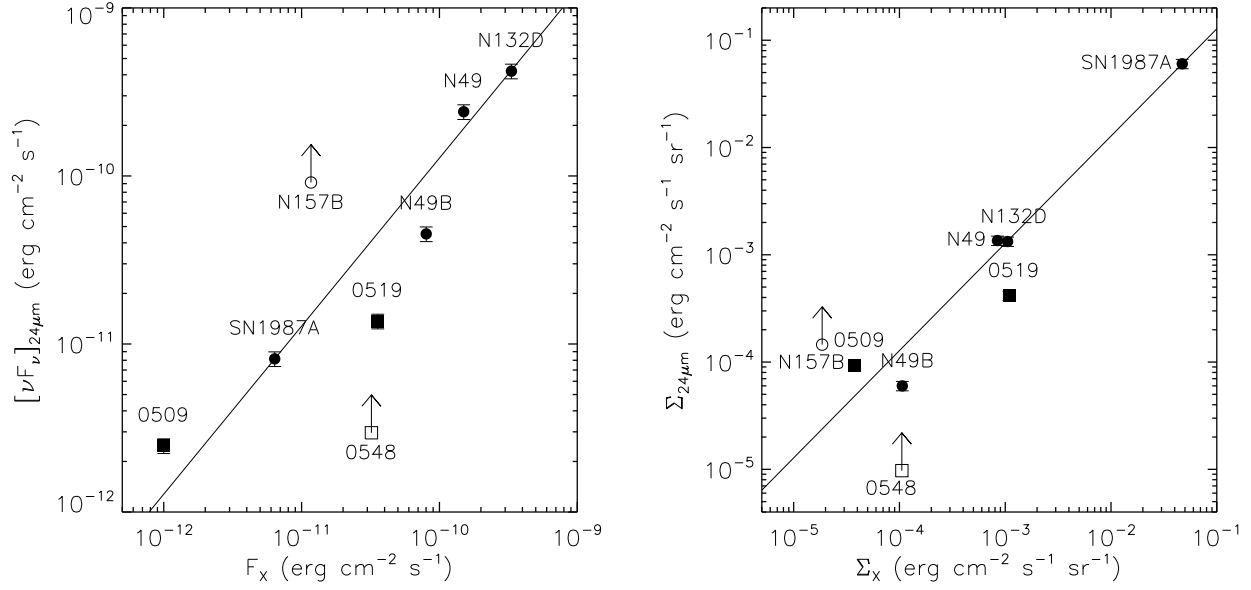

  \begin{center}
    \FigureFile(170mm,100mm){Xsvs24.eps}
    \end{center}
  \caption{AKARI 24~\micron~versus Chandra X-ray (0.3--2.1 keV) fluxes (left) and surface brightness of the SNRs (right). The solid line represents best-fit linear-regression line.}
  \label{fig:x_ir}
\end{figure}

Figure~\ref{fig:x_ir} compares the AKARI {\it L24} fluxes ($\nu F_\nu$) to the X-ray fluxes which are interstellar-absorption corrected X-ray fluxes in energy range of 0.3 to 2.1 keV. For all MIR detected SNRs, X-ray observations using Chandra are available. To have a homogeneous set of X-ray fluxes for these SNRs, we derive X-ray fluxes of these sources from archival Chandra data. For the intrinsic flux of the source, correction of the interstellar absorption is required and the resulting X-ray flux can be sensitive to assumed hydrogen column density ($N_H$). When available, we adopt $N_{H}$ from literatures, and the X-ray flux in the given energy range is estimated by fitting the X-ray spectrum from the archival data. When $N_H$ is not readily available from literatures, $N_H$ was also derived by fitting the X-ray spectrum. The uncertainty of the intrinsic flux is dominated by uncertainty in $N_H$. We consider our absorption corrected X-ray fluxes will be uncertain by factor of a few at most. In the case of SN 1987A of which time variation in X-ray flux is considerably large, we interpolate the latest fluxes from \citet{park07} to obtain the X-ray flux corresponding to the AKARI flux. Using the MIR-L and the X-ray data, it is found that there is a good correlation between the two fluxes. Both the {\it L24} and {\it L15} fluxes show good correlation with the X-ray fluxes (correlation coefficient = 0.98 and 0.90, respectively). The relation between the {\it L24} and X-ray fluxes derived from a linear least-squares fit is
\begin{equation}
   [\nu F_\nu]_{24\micron} = (1.28 \pm 0.10) \times F_X 
   \label{eq:xray}
\end{equation}
where $[\nu F_\nu]_{24\micron}$ (erg cm$^{-2}$ s$^{-1}$) is the flux in the {\it L24} band and $F_X$  (erg cm$^{-2}$ s$^{-1}$) is the X-ray flux in 0.3--2.1 keV band. For this and later fits, we do not include N157B and 0548--70.4 of which fluxes were extracted from the limited areas.

The correlation between the two fluxes is expected to some degree because the IR brightness distributions of the identified SNRs are in general correlated with their X-ray distributions except N157B. If the MIR emission is dominated by thermal dust continuum emission, the dust grains emitting the MIR emission are heated by collisions with electrons in the X-ray emitting plasma so that the two are physically associated, although the flux ratio depends on plasma temperature and therefore on the age of the remnant (e.g.,~\cite{dwek08}). The correlation in figure~\ref{fig:x_ir} (left) is {\em not} the result of both quantities being proportional to the SNR area. It is shown in figure~\ref{fig:x_ir} (right) displaying the surface brightness of the SNRs at 24 \micron~and X-ray.
There is a good correlation between the surface brightnesses too, and the best fit is given as
\begin{equation}
   \Sigma_{24\micron} = (1.28 \pm 0.01) \times \Sigma_X
   \label{eq:sbxray}
\end{equation}
where $\Sigma_{24\micron}$ $\equiv [\nu F_\nu]_{24\micron}/\Delta \Omega_S$ and $\Sigma_X$ $\equiv F_X/\Delta \Omega_S$ where $\Delta \Omega_S$ is the total solid angle of the SNR from table \ref{tab:lmcsnr}. Since the fluxes of N157B and 0548--70.4 were derived from limited areas, their surface brightnesses are lower limits. When the MIR emission originates mainly from dust, the dust temperature depends on the density and temperature of electrons. The X-ray flux is proportional to square of electron density, and at sufficiently high temperature and high densities, the dust temperature becomes only dependent on the electron density (\cite{dwek08}; see \S~5.1). This suggests the MIR surface brightness is closely related to the electron density so that the relationship can lead to the good correlation shown in figure~\ref{fig:x_ir}.

\begin{figure}
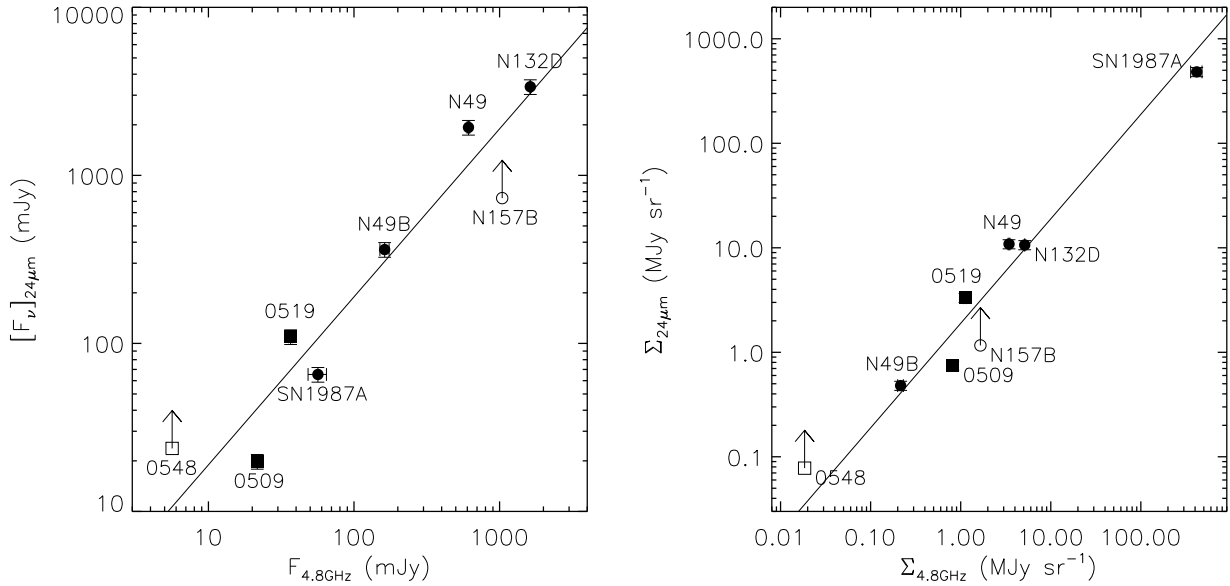

  \begin{center}
    \FigureFile(170mm,100mm){RavsL24.eps}
    \end{center}
  \caption{AKARI 24 \micron~versus 4.8 GHz radio continuum fluxes (left) and surface brightness (right). The solid line represents best-fit linear regression line.}
  \label{fig:ra_ir}
\end{figure}

\subsection{IR vs. Radio}\label{sec:irra}

Figure~\ref{fig:ra_ir} compares the AKARI {\it L24} flux ($F_\nu$) to the radio flux at 4.8 GHz. We derived the radio fluxes by using the ATCA 4.8 GHz continuum images of \citet{dickel05}. We estimated the background intensity using an annulus for most sources. The statistical errors ($1~\sigma$) of the SNRs are less than 5\%. For SN 1987A, due to the time-variability of its radio flux, we applied a flux density gradient (17.5 $\mu$Jy/day, \cite{manch02}) to the observed 4.8 GHz flux (33 mJy in day 5846) in order to obtain the radio flux corresponding to the AKARI flux, i.e. 33 mJy + 17.5 $\mu$Jy/day $\times$~($7190-5846$~day) $\simeq$ 56.5 mJy. The scaled flux is used for a further analysis. Figure~\ref{fig:ra_ir} shows that there is a fairly tight correlation between the two fluxes. The correlation is as good as that between the {\it L24} and X-ray fluxes (correlation coefficient = 0.98). The relations derived from a least-squares fit are
\begin{equation}
   [F_\nu]_{24\micron} = (1.89 \pm 0.09) \times F_{4.8 {\rm GHz}} \label{eq:radio}
\end{equation}
and
\begin{equation}
   \Sigma_{24\micron} = (1.89 \pm 0.09) \times \Sigma_{4.8{\rm GHz}}
   \label{eq:sbradio}
\end{equation}
where $F_{4.8{\rm GHz}}$ (erg cm$^{-2}$ s$^{-1}$ Hz$^{-1}$) is the radio flux at 4.8 GHz and $\Sigma_{4.8{\rm GHz}}\equiv F_{4.8{\rm GHz}}/\Delta\Omega_S$ is the radio surface brightness at 4.8 GHz. 

The correlation between the {\it L24} and radio fluxes in figure \ref{fig:ra_ir}~is remarkable considering that there is no direct physical connection between the IR and the radio synchrotron emission. The synchrotron emission is due to the compression of electrons and magnetic fields in the ISM except the pulsar wind nebula N157B. The contribution of synchrotron emission to the observed 24 \micron~emission should be negligible. Then the apparent correlation between the two fluxes in figure~\ref{fig:ra_ir} could be due to their common dependence on the physical parameters of SNRs such as shock velocity and ambient density. We note that Type Ia SNRs are faint in both IR and radio (and X-ray too), while the SNRs probably interacting with nearby molecular clouds are bright in both bands. Since Type Ia SNRs are expected to be in a lower-density environment, this trend indicates that one of the most important factors for the detection in the IR is likely to be the density of the ambient medium. 

The correlation in figure~\ref{fig:ra_ir} would be compared to that among galaxies. It has been known for over three decades that there is a tight correlation between the far-infared and radio emission from galaxies, and recent Spitzer studies have showed that even MIR and radio has a fairly good correlation ($F_{24\micron}/F_{1.4{\rm GHz}}~\sim~7-25$, \cite{apple04, boyle07}). For comparison, the corelation in figure~\ref{fig:ra_ir} implies the 24 \micron~to 1.4 GHz flux ratios of 1.0 using a spectral index of $-0.5$ for radio synchrotron emission. Therefore, if the 1.4 GHz radio emission from galaxies is mainly from SNRs, the contribution of SNRs to the 24 \micron~fluxes from galaxies would be $4-14$~\%, the rest of which might be contributed from the star-formation activity.

\section{Discussion}
\subsection{Origin of Infrared Emission}\label{sec:irorigin}

There are four sources of the MIR ($10-30$~\micron) emission in SNRs in general; ionic and/or molecular lines, thermal dust continuum emission, PAH bands, and non-thermal synchrotron emission (\cite{koo07} and references therein). Synchrotron emission is usually negligible except for young PWN such as Crab Nabula. PAH emission has not been detected towards SNRs except N132D where it appears as a $15-20$ \micron~emission hump superposed on a strong dust continuum \citep{tappe06}. It is usually either the thermal emission from collisionally-heated dust grains in hot plasma or the forbidden lines from the elements such as Ne, O, Fe ions and pure rotational H$_2$ lines that dominate the emission in this wave band. In order to distinguish between line- and dust-dominated SNRs, we can compare an IR color of a SNR to a theoretical prediction from the main emission mechanisms in a color-color diagram. Besides this, comparison of IR morphology to X-ray and optical can be an another useful means. When IR morphology is very similar to that of X-ray but optical, this might indicate that thermal dust emission is dominant in a SNR because, if the line radiation from a radiative shock is dominant, the IR morphology would be similar to the optical. However, in case of a Balmer-dominated  SNR, resemblance between IR and optical often looks as good as that between IR and X-ray so that other physical conditions such as an age should be considered, too. 

For three Type II SNRs, N132D, SN 1987A, and N49, spectroscopic observations have been done using Spitzer. It is shown that the MIR spectra ($10-30~\micron$) of the former two SNRs are dominated by thermal dust continuum with small contribution from several ionic lines and PAH emission \citep{tappe06, bouchet06, dwek08}. Their AKARI MIR colors are consistent with these spectroscopic results (see next). On the other hand, the MIR spectrum of N49 is found to be dominated by ionic lines from shocked gas without any substantial dust continuum emission \citep{williams06}. (There are some pure rotational H$_2$ lines too, but their contribution is relatively small.) This is interesting because the AKARI MIR colors of N49 is aligned close to those of thermal dust emission (figure~\ref{fig:snrccd}). We check the possibility that the MIR spectra of the positions where the spectroscopic observations were performed do not represent the spectrum of the entire SNR in N49. The spectroscopic observations were performed toward two positions, and their synthetic IRC colors are ({\it L15}/{\it L24}, {\it S11}/{\it L15})=(0.41, 0.26) and (0.23, 0.28), respectively. For comparison, the observed IRC color of the entire SNR N49 is $\sim (0.45, 0.38)$, which is not significantly different from the synthetic ratio\footnote{The referee pointed out that, according to her/his recent spectral mapping of N49, there are faint signs of continuum emission in some remnant regions, but line emission is a strong contributor in all IR-bright areas of this SNR. This supports our argument.}. The somewhat higher {\it S11}/{\it L15} ratio could be either due to variations of line intensities over the remnant or possibly due to dust emission. The important point is that {\em the SNRs dominated by ionic line emission can have colors similar to modified blackbodies in the ({\it L15}/{\it L24}, {\it S11}/{\it L15}) diagram}. Important lines in these bands are [Ne II] $12.81~\micron$ ({\it S11}), [Ne III] $15.56~\micron$ ({\it L15}), [O IV] $25.90~\micron$ and [Fe II] $26.00~\micron$ ({\it L24}) lines. RCW 103 is far off from the modified blackbody line apparently because its Ne lines are much stronger than O or Fe lines in contrast to N49. 

For the other five SNRs, no MIR spectroscopic data are available at the moment. For 0509--67.5 and 0519--69.0, Spitzer recently obtained their spectra which have not been released yet, and an IRS observation of N157B is also planned. 0509--67.5 and 0519--69.0 are young, Balmer-dominated Type Ia SNRs. Their shock is very fast ($\gtrsim 3,000$~km/s) and non-radiative \citep{ghava07}, so that we do not expect strong IR ionic or molecular line emission. Their MIR emission might be from shock-heated dust grains. The other Type Ia SNR, 0548--70.4, is also Balmer-dominated, but significantly older ($\sim 7\times 10^3$~yr) \citep{hendrick03}. An interesting feature of this remnant is the X-ray emitting gas in the central region, where the bright MIR emission is detected by AKARI (figure \ref{fig:mirsnr2}). It shows enhanced metal abundance and could be SN ejecta swept-up by reverse shock \citep{hendrick03}. On the other hand, there are bright H$\alpha$/[O III]-emitting clumps mixed with X-ray emitting gas, and they could be either dense ejecta or interstellar clumps swept up by slow, radiative shocks (cf. \cite{ghava07}). The brightness distribution of the MIR emission is considerably different from the X-rays. There are several bright clumps seen in the X-rays, and some of them are extended towards the south where there is no distinct MIR emission. Instead, the optical knots in the H$\alpha$ images are mainly distributed inside of the central MIR emitting region. This remnant is also considerably off from the thermal dust-emission line in figure \ref{fig:snrccd}. This suggests that the MIR emission in the central area is likely dominated by ionic lines from radiative shocks rather than by continuum emission from hot dusts.

Two remaining SNRs are N49B and N157B, both of which are Type IIs. N49B is a middle-aged shell-type SNR, and its MIR morphology is very similar to the X-rays whereas the H$\alpha$ image hardly shows any shell structure \citep{williams99}. This suggests that the MIR emission is likely dominated by thermal dust emission. But the bright portion of the southern shell and also the clump in the eastern part appear bright in H$\alpha$/[O III] emission \citep{mathewson83}, so that there could be some contribution from ionic line emission. N157B is a Crab-like SNR. As we described in \S~3.1, we have not detected any appreciable IR emission corresponding to the PWN, but detected IR emission corresponding to the H$\alpha$-emitting nebula in the east of the PWN both in MIR and NIR bands (figure \ref{fig:mirsnr2}$-$\ref{fig:nirsnr}; cf. \cite{williams06}). The MIR emission is probably dominated by ionic lines. In summary, among 5 SNRs, the MIR emission of 0509--67.5, 0519--69.0, and N49B are thought to be dominated by thermal dust continuum, while 0548--70.4 and N157B are by ionic lines. 

For SNRs that are considered to be dominated by thermal dust emission, we fit their AKARI {\it S11}, {\it L15}, and {\it L24} fluxes by a single-temperature dust emission and the results are summarized in table \ref{tab:sedfit}. For this calculation, we first color-corrected the measured fluxes by assuming the modified blackbody curve (equation \ref{eq:modbb}). The correction factors range 0.8 -- 1.1, 1.3 -- 2.3, and 0.9 -- 1.1 for the {\it S11}, {\it L15}, and {\it L24}, respectively. Note that the correction factors of the {\it L15} band are large because the dust mass absorption coefficient, $\kappa_\nu$, in the assumed model has a dip near 15 \micron~owing to the characteristic of the silicate. The derived temperatures vary from 86 to 185 K, which corresponds to the luminosity and mass ranges of $0.8-140 \times 10^{36}$~erg s$^{-1}$ and $0.1-130 \times 10^{-4}M_{\odot}$. The derived temperatures agree with the results derived from the Spitzer spectroscopy on the N132D and SN 1987A. For N132D, \citet{tappe06} applied a two-component fit to its Spitzer IRS spectrum of the southeastern rim to derive temperatures of 58 and 110 K. The higher temperature component contributes most of the emission at $\lesssim 30$~\micron. Our temperature (96 K) is somewhat lower than their 110 K, but this could be due to the lower-temperature component. For SN 1987A, \citet{bouchet06} and \citet{dwek08} showed that its Spitzer IRS spectrum is well described by a single-temperature thermal dust emission at $\sim 180$~K. 

Dust grains in hot plasma are mainly heated by collisions with electrons. Dust temperature depends on the electron temperature and density. When the plasma temperature is sufficiently high, most electrons go through dust grains and the dust temperature becomes dependent only on electron density \citep{dwek08}. \citet{dwek08} showed that $0.023-0.22$~$\micron$-sized dust grains in the hot plasma at $3.5\times 10^6$~K with a density of $(0.3-1)\times 10^4$ cm$^{-3}$ can reach 180 K, the temperature observed in SN 1987A. The dust temperatures of the other four SNRs are $86-99$~K. Such temperatures can be achieved for $0.01-0.1~\micron$-sized dust grains in the hot plasma at $\sim 10^7$~K when the electron density is higher than $10-20$~cm$^{-3}$. We note that for Type II SNRs (N132D and N49), X-ray observations indicated such high densities \citep{williams06,park03}. For two Type Ia SNRs (0509--67.5 and 0519--69.0), an analysis of H$\alpha$ lines yielded lower limits on electron densities, 1.6 and 7.7 cm$^{-3}$ \citep{borko06}. The MIR emissions in these remnants are mostly from confined regions, and it is possible that these regions are where the blast wave is propagating into dense interstellar material of higher electron density.

\begin{figure}
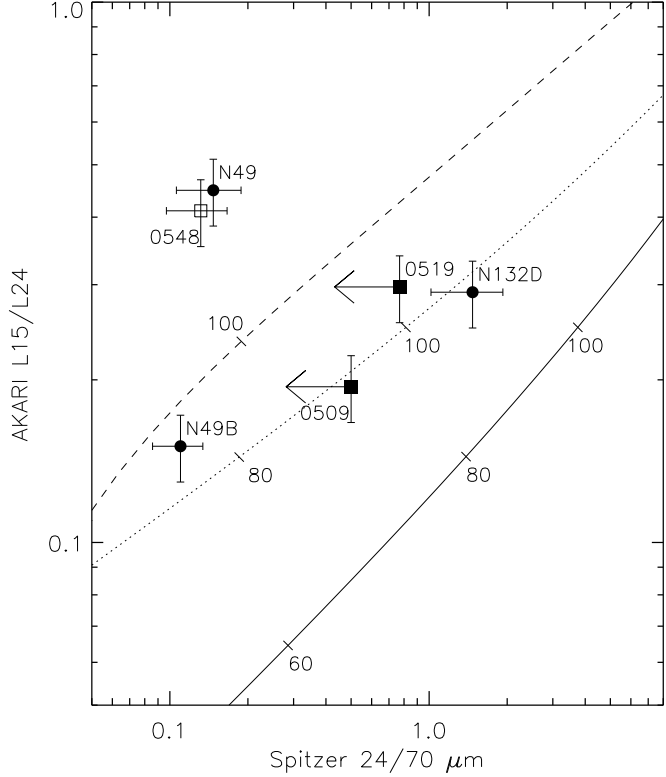

  \begin{center}
    \FigureFile(100mm,100mm){lmcsnrMirCCD.eps}
    \end{center}
  \caption{AKARI {\it L15}/{\it L24} flux ratio versus Spitzer 24/70~\micron~flux ratio. Solid line represents the expected ratios from modified blackbody curve of thermal dust emission with a single component. Dotted and dashed lines show the ratios of two-component dust emission with T$_{cold}=30$~and 40 K, respectively when $M_{cold}/M_{warm}=500$ (see text for detail).}
  \label{fig:mirccd}
\end{figure}

\subsection{Dust Destruction by SNR Shocks}\label{sec:dust}

SNR shocks are the main source of dust destruction. Dust grains are destroyed by sputtering and grain-grain collisions behind the shock \citep{jones04}. Recently, \citet{borko06} and \citet{b_willi06} observed Type Ia and core-collapse SNRs using Spitzer, respectively. They found that the dust destruction by sputtering is necessary to reproduce the observed 70/24 \micron~ratios according to their shock model calculation. They concluded that about 40 \% of the mass in dust grains has been destroyed and smaller grains ($\le0.04~\micron$) has been shattered up to 90 \% in both types of SNRs. 

We note that the dust temperatures derived from the AKARI {\it S11}, {\it L15}, and {\it L24} fluxes are considerably higher than those of \citet{borko06} and \citet{b_willi06}. If we calculate the 70/24 \micron~flux ratios using our dust temperature, the ratios are significantly smaller than the Spitzer results. This is shown in figure \ref{fig:mirccd}~which is a color-color diagram of AKARI {\it L15}/{\it L24} versus Spitzer 24/70 \micron. Spitzer 24/70 \micron~ratios are adopted from literatures \citep{borko06, b_willi06, williams06}. It is clear that a single temperature cannot describe the full spectral energy distribution (SED) of these SNRs including the ones probably dominated by thermal dust emission. It is interesting that N49 and 0548--70.4, two SNRs thought to be line-dominated, deviate more largely from the dust-continuum curve. From previous studies, it is known that the SED of SNRs usually requires two components, cold and warm dusts. We can also explain the color of SNRs in figure 9 with an extra cold component at $20-40$ K and with the mass ratio to the warm component of 
$50-1000$. The lines in figure 9 show the cases when the temperature of the cold component is 30 and 40 K and the mass ratio is 500. The colors of the line-dominated SNRs, however, cannot be easily described by an extra cold component. We can fit their colors using a cold component ($30-40$ K) plus a warm component ($100-150$ K) with a mass ratio of $\sim10000$. But, since both SNRs are middle-aged, the temperature of the warm component appears too high for them. This supports our conclusion that these SNRs are line-dominated. The strong ionic lines around 70 $\micron$ are [O I] 63 $\micron$~and [O III] 88 $\micron$~which might have led the 24/70 $\micron$~ratio to decrease. This might indicate that the discrepancy is to some extent due to the line emission.

Multiple dust temperatures could happen if the ambient ISM is clumpy, so that the shock velocities and therefore dust temperatures differ for the clump and interclump medium as suggested by \citet{tappe06} in N132D. It can also happen if the grains size distribution does not follow a simple power law. In this regard, it is worth to note that previous IRAS studies showed that the IR emission from SNRs usually require two components; a cold component dominating $\gtrsim 70$~\micron~and a warm component dominating $\lesssim 25$~\micron~\citep{arendt89,saken92}. Figure \ref{fig:mirccd} is consistent with previous IRAS results, and suggests that the 70 \micron~fluxes in these SNRs might be dominated by cold dust component. This result implies that the destruction rate of grains in these SNRs may not be as high as previously estimated from the 24/70 \micron~flux ratios.

\section{Summary}

We performed an IR study of 21 LMC SNRs listed in table~\ref{tab:lmcsnr} using the AKARI IRC survey data. We systematically examined the AKARI images in 3 $\micron$~({\it N3}), 7 $\micron$~({\it S7}), 11 $\micron$~({\it S11}), 15 $\micron$~({\it L15}), and 24 $\micron$~({\it L24}) bands, and detected eight SNRs with associated IR emission (table~\ref{tab:ircflux}). In the {\it S11} band, we could reveal the SNR features much more clearly by subtracting the background emission using the {\it S7} band image. We summarize our main results below: 

1.	In all eight sources, we can see distinct MIR emission features at the position of SNRs in the {\it S11}, {\it L15}, and {\it L24} bands. Our {\it L24} images are consistent with their Spitzer 24 $\micron$~images. The {\it S11} and {\it L15} images are new in these wavebands, and show that individual SNRs have compatible morphologies in three bands. This suggests that the MIR emissions in these three bands are of the same origin. In shorter wavebands, three SNRs are clearly visible: N49, SN 1987A, and N157B. 

2.	In the {\it S11}/{\it L15} flux ratio versus the {\it L15}/{\it L24} flux ratio diagram, the SNRs are well-aligned along a track of thermal dust emission. Type Ia SNRs are located in the lower left with low ratios, while Type II SNRs are spread from low to high ratios. SN 1987A has the highest ratios. The alignment along the dust emission line does not necessarily imply that the origin of the MIR emission is thermal dust continuum because the SNRs dominated by ionic lines can have colors similar to modified blackbodies in this diagram. But the diagram is still helpful to distinguish between line-dominated and dust-dominated SNRs, taking account of their physical contexts.

3.	The flux in the {\it L24} band has a good correlation with both the soft X-ray and radio fluxes. There is also a good correlation between the corresponding surface brightnesses. The correlation between the MIR and radio is remarkable, considering that there is no direct physical connection between the two. The correlations might be due to their common dependence on the physical parameters of SNRs, particularly the density of the ambient medium. The correlation yields $\sim1.0$ for the ratio of 24 \micron~to 20 cm radio fluxes of SNRs, and it implies that the contribution of SNRs to the 24 \micron~fluxes from galaxies would be $4\%-14$\% at most.

4.	We consider that the MIR emission from the five SNRs is dominated by thermal dust emission: two Type Ia SNRs (0509--67.5 and 0519--69.0) and three Type II SNRs (N132D, N49B, and SN 1987A). We derived dust temperatures of $\sim90-190$~K from their $10-25~\micron$~fluxes. Their {\it L15}/{\it L24} color temperatures are generally higher than the Spitzer 24/70 \micron~color temperatures, which suggests that the dust emission model at single equilibrium temperature cannot describe the full SED of these SNRs. This also implies that understanding of the full SED is essential for estimating the dust destruction rate of grains by SNR shocks.

\section*{Acknowledgements}
This work is based on observations with AKARI, a JAXA project with the participation of ESA. We wish to thank all the members of the AKARI project. We thank the anonymous referee for helpful comments, which have considerably improved the paper. We also would like to thank Snezana Stanimirovi\`c for useful comments on the IR/radio correlation of galaxies. This work was supported by the Korea Science and Engineering Foundation (R01-2007-000-20336-0) and the Korea Research Foundation (R14-2002-058-01003-0).

\newpage

\begin{table}
  \caption{SNRs in the AKARI LMC Survey} \label{tab:lmcsnr}
  \begin{center}
   \begin{tabular}{lccccccc}
    \hline
   & R.A. & Dec. & Size & 4.8 GHz Flux &  & AKARI & AKARI  \\ 
~~~SNR & (J2000) & (J2000) & (arcmin) & (mJy) & Type & Coverage & Detection \\
~~~~(1) & (2) & (3) & (4) & (5) & (6) & (7) & (8) \\
   \hline
   \hline
0509--67.5     & 05:09:31  & -67:31:17 & 0.56$\times$0.56 & 22 & Ia & A & {\it S11}/{\it L15}/{\it L24}  \\ 
0513--69.2      & 05:13:14  &   -69:12:20  & 4.8$\times$3.4  & 78 & I? & B2 &\dots \\ 
0519--69.0      & 05:19:35  &   -69:02:09 & 0.67$\times$0.58& 37  & Ia & A & {\it S11}/{\it L15}/{\it L24} \\ 
0520--69.4      & 05:19:44  &   -69:26:08 &3.1$\times$2.7 & 28 & ?& C &\dots \\ 
SNR in N44    & 05:23:07  &   -67:53:12 &4.1$\times$4.4 & $<$648~~~ &?& A & \dots\\ 
DEM L175A     & 05:24:20  &   -66:24:23 & 4.4$\times$3.4 & $<$133~~~ &I? & A& \dots \\ 
N132D         & 05:25:04  &   -69:38:20 &2.2$\times$1.7 & 1625~~ & II & B1 & {\it S11}/{\it L15}/{\it L24} \\ 
N49B          &  05:25:25 &    -65:59:19 & 2.7$\times$2.3 & 162~ & II & A & {\it S11}/{\it L15}/{\it L24} \\ 
N49   	       &  05:26:00 &    -66:04:57 & 1.5$\times$1.4 & 611~ & II & A & all bands \\ 
0528--69.2      &  05:27:39 &    -69:12:04 & 2.7$\times$2.2& 16 &  II? & A & \dots \\ 
DEM L204      &  05:27:54 &    -65:49:38 & 5.2$\times$4.9& 38 & II? & B1 & \dots\\ 
0534--69.9      &  05:34:02 &    -69:55:03 & 2.6$\times$2.2& 26 & Ia? & A & \dots \\ 
SN 1987A      &  05:35:28 &    -69:16:11 & 0.04$\times$0.04 & $33^{a}$ & II & A & all bands \\ 
Honeycomb     & 05:35:46  &   -69:18:02  &  2.2$\times$1.2 & 10& ? & A & \dots \\ 
DEM L249      &  05:36:07 &    -70:38:37 &3.1$\times$2.3& 9 & Ia? & C &  \dots \\ 
N157B	        & 05:37:49  &   -69:10:20  & 3.1$\times$2.4 & 1042~~ & II(Crab-like) & A& {\it N3}/{\it S11}/{\it L15}/{\it L24} \\ 
N158A	        & 05:40:11  &   -69:19:55  & 1.3$\times$1.1& 341~ & II(Crab-like) & A &\dots \\ 
SNR in N159   & 05:39:59  &   -69:44:02  & 2.0$\times$1.8& 1214~~ & ? & A & \dots\\ 
DEM L316B     & 05:46:59  &   -69:42:50  & 3.4$\times$2.8& $317^{b}$~ & II & A & \dots \\ 
DEM L316A     & 05:47:22  &   -69:41:26  & 2.2$\times$1.9& $317^{b}$~ & Ia? & A & \dots \\ 
0548--70.4      & 05:47:49  &   -70:24:54  & 2.0$\times$1.8& 6 & Ia & A& {\it S11}/{\it L15}/{\it L24}\\ 
   \hline
   \end{tabular}
   \end{center}
\noindent Notes.--- Col. (1)--(4): SNR names, positions, and angular sizes from the MC SNR Atlas by R. Wiliiams (http://www.astro.uiuc.edu/projects/atlas/). Angular sizes are mostly in optical. The size of SN 1987A is a radius of the inner equatorial ring \citep{bouchet06}. Col. (5): 4.8 GHz flux estimated from the radio image of  \citet{dickel05} available at the NCSA Astronomy Digital Image Library. The statistical errors of SNRs are $\lesssim$~5\%, and limits are 3 $\sigma$. Col. (6): SNR type from mostly \citet{williams99} and other literatures. Except the two Crab-like SNRs, Type II SNRs are shell type SNRs of core-collapse SN origin. For unclear/undefined SNR types, question marks are used. Col. (7): AKARI coverage states how completely the AKARI IRC LMC survey includes the SNR area in the IRC bands. A: all covered by AKARI in all bands, B1: wholly in the MIR-L and partially in the NIR/MIR-S, B2 : vice versa, and C: only partially in some bands. Col. (8): AKARI bands in which the IR emission from the SNR is detected.  \\
\noindent $^{a}$ Observed at Feb. 14, 2003 (day 5846 after the SN explosion). A scaled flux to the day 7190 (56.5 mJy) is used for a further analysis (see \S~4.2). 

$^{b}$ Since DEM L316B and DEM L316A are located very closely, the total radio flux of the two SNRs is estimated.
 \end{table}

\newpage

\begin{table}
  \caption{AKARI IRC Flux and Color Estimates}\label{tab:ircflux}
 \begin{center}
  \begin{tabular}{lccccccc}
\hline
   & {\it N3} & {\it S7} & {\it S11} &{\it L15} & {\it L24} &&  \\ 
SNR & (mJy) & (mJy) & (mJy) &(mJy) & (mJy) & {\it S11}/{\it L15}& {\it L15}/{\it L24} \\
  \hline
  \hline
0509--67.5 
	&$<$0.1&$<$0.2&0.3 $\pm$ 0.1 &~~3.9 $\pm$ 0.4&20 $\pm$ 2
	&0.08 $\pm$ 0.03&0.19 $\pm$ 0.03\\ 
0519--69.0 
	&$<$0.6&$<$1.3&2.5 $\pm$ 0.3&33 $\pm$ 3~&110 $\pm$ 11
	&0.08 $\pm$ 0.01& 0.30 $\pm$ 0.04\\ 
$\mathrm{N132D}^{a}$ 
  	&\ldots&\ldots&60 $\pm$ 6~~&980 $\pm$ 98&3370 $\pm$ 340
   	&0.07 $\pm$ 0.01&0.26 $\pm$ 0.04\\
N49B 
	&$<$0.6&~$<$1.6&3.5 $\pm$ 0.4&55 $\pm$ 5&360 $\pm$ 36
	&0.06 $\pm$ 0.01&0.15 $\pm$ 0.02\\ 
N49  
	& 36 $\pm$ 4~ & 280 $\pm$ 28~ &330 $\pm$ 33~~&870 $\pm$ 87
	&1930 $\pm$ 193&0.38 $\pm$ 0.05&0.45 $\pm$ 0.06\\
SN 1987A 
	&1.5 $\pm$ 0.1 &4.9 $\pm$ 0.5&32 $\pm$ 3~~&43 $\pm$ 4&65 $\pm$ 7
	&0.75 $\pm$ 0.11&0.66 $\pm$ 0.09\\ 
$\mathrm{N157B}^{b}$ 
	&4.3 $\pm$ 0.4&$<$1.9&39 $\pm$ 4~~&280 $\pm$ 28
	&730 $\pm$ 73&0.14 $\pm$ 0.02&0.38 $\pm$ 0.05\\ 
0548--70.4$^{b}$ 
	&$<$0.2&$<$0.6&0.6 $\pm$ 0.2&10 $\pm$ 1&24 $\pm$ 2
	&0.06 $\pm$ 0.02&0.41 $\pm$ 0.06\\ 
	
\hline
  \end{tabular}
 \end{center}
\noindent Notes.--- Fluxes at reference wavelengths of each band are given. The reference wavelengths are 3.2, 7.0, 11.0, 15.0, and 24.0 \micron~for the {\it N3}, {\it S7}, {\it S11}, {\it L15}, and {\it L24}, respectively. {\it S11} fluxes are from {\it S11}-{\it S7} images except N49 and SN 1987A. Fluxes are not color-corrected. Flux errors are 1 $\sigma$ and limits are 3 $\sigma$. The last two columns are color ratios of the {\it S11}/{\it L15} and {\it L15}/{\it L24}. \\
\noindent $^{a}$~N132D has not been fully covered in the {\it S11} band. We derive the {\it S11} flux by first deriving the ratio of the {\it S11} to {\it L15} flux from the shared area and then by multiplying it to the total {\it L15} flux.\\
\noindent $^{b}$~The fluxes of N157B and 0548--70.4 are not from the entire SNRs but from the limited areas (figure \ref{fig:mirsnr2}).    
\end{table}


\begin{table}
  \caption{Dust Properties of Five SNRs probably dominated by Thermal Dust Emission}\label{tab:sedfit}
 \begin{center}
  \begin{tabular}{lccc}
 \hline
  & T(dust)&$L_{IR}$& Dust Mass\\
SNR & (K)&(erg s$^{-1}$)& ($M_{\odot}$)\\
 \hline
 \hline
 
0509--67.5 & 94 $\pm$ 3& 8.1 $\times~10^{35}$ & 8.7 $\pm$ 2.5 $\times~10^{-5}$ \\
0519--69.0 & 99 $\pm$ 4 & 4.5 $\times~10^{36}$ & 3.6 $\pm$ 1.0 $\times~10^{-4}$\\
N132D	   & 96 $\pm$ 4& 1.4 $\times~10^{38}$& 1.3 $\pm$ 0.3 $\times~10^{-2}$ \\
N49B & 86 $\pm$ 3 & 1.6 $\times~10^{37}$& 2.8 $\pm$ 0.7 $\times~10^{-3}$\\
SN 1987A& 185 $\pm$ 15 & 3.6 $\times~10^{36}$ & 1.0 $\pm$ 0.2 $\times~10^{-5}$ \\

 \hline
  \end{tabular}
 \end{center}   
\end{table}

\newpage

\appendix

\section*{Brief Description on Individual SNRs}

\noindent \textbf{0509--67.5} --- This Balmer-dominated remnant is one of the youngest SNRs in the LMC, and its X-ray spectra indicate that it is originated from Type Ia SN explosion \citep{hughes95}. There is a bright, elongated feature in the southwest of the SNR in the MIR-L bands, and it coincides with the brightest portion of the shell structure seen in both the X-ray and optical images \citep{borko06}. {\it S11} and {\it S11}-{\it S7}~images also show emission at the same position. Because it is faint, the morphologies in the {\it S11} and {\it S11}-{\it S7} are not clear. In addition to the bright southwestern limb, in the  {\it L24}, the faint emission from the rest of the SNR is visible as seen in the X-ray and optical. The bright point-like source just outside the southwestern boundary of the contour in the {\it S11} is a background source (2MASSJ 05092882-6731307).  

\noindent \textbf{0519--69.0} ---  This is another young Balmer-dominated SNR probably originated from Type Ia SN explosion \citep{hughes95, ghava07}. The shell structure well-defined in H$\alpha$~and X-ray is clearly seen in the MIR-L bands. Also, it shows three bright knots in the north (N), east (E), and southwest (SW) along the limb. While only the N and the E knots are visible in the {\it S11} image, all three knots are clearly visible in the {\it S11}-{\it S7} image with the morphology nearly identical to those of the MIR-L. The N and the E knots spatially correspond to the relatively bright region in X-ray and  H$\alpha$, whereas the SW knot does not have specific counterpart in those bands. This discrepancy of the knots indicates the different characteristics among them. 

\noindent \textbf{N132D} --- This SNR belongs to young oxygen-rich SNRs that are the product of the core-collapse SNe \citep{morse95}. The remnant is one of the brightest LMC SNRs and shows a well-defined shell structure in the MIR bands as in the X-rays. The southeastern (SE) rim contains enhanced IR emission which might be caused by the interaction with a molecular cloud in this area \citep{banas97, tappe06}. Only the bright SE shell is visible in the {\it S11} band together with the northwestern knot named ``West Complex'' by the previous optical observation \citep{tappe06}. However, the {\it S11}-{\it S7} image shows the morphology just same as that of the MIR-L bands. The central bright emission in the {\it S11} band, showing weak correlation with X-ray, might be mainly attributed to background emission. The IR morphology is somewhat different from that of the optical which shows relatively fainter shell compared to the bright ejecta region at the center \citep{borko07}. Using the Spitzer IRS observations, \citet{tappe06} revealed the dominant dust continuum with the first detected PAH emission. 

\noindent \textbf{N49B} --- This is a middle-aged SNR of core-collapse SN origin \citep{hughes98, park03}. The SNR shell is clearly seen in the MIR-L bands, and its morphology is similar to that in X-ray. The SNR has several patchy emission along the limb including the particularly bright southern rim in the MIR-L bands. There is some diffuse IR emission in the inner region with a belt-like feature crossing the shell. Even though the {\it S11} image does not show prominent emission related to the SNR, the {\it S11}-{\it S7} image reveals some emission such as the southern rim, the belt-like feature, and the northwest shell similar to the features in the MIR-L bands.  

\noindent \textbf{N49} --- This Type II SNR is unique in showing very strong emission in all IRC bands. Previous X-rays and radio observations show fairly clear emission over the entire shell of the SNR with the peak in the east \citep{williams06}. While the IRC images also show the similar morphology, the IR peak has a different position from the X-rays/radio. Besides, the SNR has a lack of IR emission in the northwest unlike the X-rays/radio although the complete shell of the X-rays/radio is marginally detected in the MIR-L bands. A bright wedge-shaped feature in the east is clearly visible in all IRC bands, which is quite similar to that seen in the HST observations \citep{bilikova07}. Also, the overall IR morphology has good correspondence to that of optical. In the {\it S7} band, there appears a protrusion that extends to east from the wedge-shaped filament. Since the structure extends beyond the SNR boundary, it is not likely to be physically associated with the SNR. But it is worth to note that it is spatially coincident with the ambient molecular cloud possibly interacting with the SNR \citep{banas97}.

\noindent \textbf{SN 1987A} --- This newly formed SNR has just started its interaction with dense circumstellar material around the inner ring (e.g., \cite{park05}). The effect of this encounter has been detected as a rapid brightness change at various wavelengths including IR \citep{bouchet06}. AKARI observed this SNR at Oct. 31 -- Nov. 4, 2006 (day 7190--7194) and the estimated flux agrees with the IR flux variation found by the recent Spitzer study \citep{dwek08}. The remnant appears point-like in all IRC bands, but the previous observation with higher resolution shows the resemblance with X-rays rather than optical in terms of the brightness distribution (e.g., see figure 3 and 16 in \cite{bouchet06}). The ring-like feature around the SNR in the MIR-L bands is the shape of point-spread-function (PSF) of the IRC.

\noindent \textbf{N157B} --- This remnant is one of the two known Crab-like SNRs in the LMC. No appreciable IR emission related to the pulsar or PWN has been detected. The bright emission is mostly originated not from the SNR but from nearby sources including a small molecular cloud in the south \citep{johan98}, possibly undergoing star-forming activities in progress. However, it is found that there are some features such as the horseshoe-shaped one and the arm-like emission towards the northeast in all IRC bands except the {\it S7}. These features correspond well to the H$\alpha$ emission of the SNR observed by \citet{chu92}. This indicates they are associated with N157B. 

\noindent \textbf{0548--70.4} --- This middle aged ($\sim$7,000 yr), Balmer-dominated SNR has been categorized into a Type Ia remnant because its observed ratio of oxygen to iron is much lower than that from typical core-collapse SNe \citep{hendrick03}. The shell structure detected in H$\alpha$~and X-rays is not clearly seen in the IRC images while the Spitzer image shows the outer shell similar to the H$\alpha$/X-ray owing to its higher resolution \citep{borko06}. Nevertheless, there is a distinct emission in MIR bands that has a spatial correspondence to the central region with X-ray emission. While the overall shape of the IR emission is similar to that of the X-rays, its brightness distribution is quite different from the X-rays. Instead, the brightness distribution of MIR emission seems closer to the H$\alpha$ including the bright knot at the center.       

\end{document}